\documentclass[aps,pre,twocolumn,10pt,showpacs]{revtex4-1}

\usepackage{times}
\usepackage{amsmath}
\usepackage{amssymb}
\usepackage{graphicx}
\usepackage{color}
\usepackage{verbatim}

\newcommand{\eqn}  {Eq.~}

\newcommand  {\avg}[1]        {\langle#1\rangle}
\newcommand  {\avgmicro}[1]   {\langle#1\rangle_{\mu}}
\newcommand  {\avgcenter}[1]  {\langle#1\rangle_{\mathrm{c}}}

\newcommand  {\nup}           {N_{\mathord{\uparrow}}}
\newcommand  {\nb}            {N_\mathrm{b}}

\newcommand  {\ket}[1]        {\lvert#1\rangle}

\newcommand  {\bramidket}[3]  {\langle#1\vert#2\vert#3\rangle}
\newcommand  {\abs}[1]        {\lvert#1\rvert}

\begin{document}

\title{Finite-size scaling of eigenstate thermalization}

\author{W. Beugeling}
\author{R. Moessner}
\author{Masudul Haque}

\affiliation{Max-Planck-Institut f\"ur Physik komplexer Systeme, N\"othnitzer Stra\ss e 38, 01187 Dresden, Germany}

\date{\today}
\pacs{05.30.-d,05.70.Ln,75.10.Pq}


\begin{abstract}

According to the eigenstate thermalization hypothesis (ETH), even isolated quantum systems can
thermalize because the eigenstate-to-eigenstate fluctuations of typical observables vanish in the
limit of large systems. Of course, isolated systems are by nature finite, and the main way of
computing such quantities is through numerical evaluation for finite-size systems. Therefore, the
finite-size scaling of the fluctuations of eigenstate expectation values is a central aspect of the
ETH. In this work, we present numerical evidence that for generic non-integrable systems these
fluctuations scale with a universal power law $D^{-1/2}$ with the dimension $D$ of the Hilbert
space. We provide heuristic arguments, in the same spirit as the ETH, to explain this universal
result. Our results are based on the analysis of three families of models, and several observables
for each model. Each family includes integrable members, and we show how the system size where the
universal power law becomes visible is affected by the proximity to integrability. 

\end{abstract}


\maketitle

\section{Introduction}

In recent years, non-equilibrium unitary evolution of isolated quantum systems has emerged as a key
topic in many-body physics.  In this context, the issue of thermalization in isolated quantum
systems has received fresh and growing attention.  The eigenstate thermalization hypothesis (ETH) is
widely thought to encapsulate the mechanism by which thermalization occurs in isolated
non-integrable systems \cite{Deutsch1991,Srednicki1994,RigolEA2008}.

The basic statement of the ETH is that, for a large isolated system, the diagonal matrix elements of
typical observables in the Hamiltonian eigenstate basis, known as the eigenstate expectation values
(EEVs), depend smoothly on the corresponding energy eigenvalues.  Despite intense recent research
\cite{PolkovnikovEA2011,RigolEA2008,KollathEA2010,Roux2010PRA81,Motohashi2011,IkedaEA2011, GenwayEA2012,RigolSrednicki2012,KhatamiEA2012,BrandinoEA2012,Rigol2009PRA,Rigol2009PRL,SantosRigol2010,YurovskyOlshanii2011,NeuenhahnMarquardt2012,CassidyEA2011,IkedaEA2013,SteinigewegEA2013},
understanding of several aspects of the ETH remains incomplete.  For example, it is not fully known
exactly which observables will or will not serve as ``typical'' observables.  Another issue is the
specification of ``large'' isolated systems --- how large does the system have to be?  Clearly, a
proper understanding of this question requires a finite-size scaling study of the ETH.  This is an
important question for any actual experimental study of thermalization, because any isolated system
is in practice finite.  Size dependence is also vital for evaluating numerical studies, which are
performed on finite systems.  This is the subject of the present manuscript.

It is generally understood that the fluctuations ($\sigma_{\Delta{A}}$) of EEVs should decrease
exponentially with system size \cite{IkedaEA2011, SteinigewegEA2013, RigolSrednicki2012,
  Deutsch1991}, so that the EEVs become very smooth as a function of energy for reasonably large
isolated systems.  For discrete systems with a finite Hilbert space, this means a power-law
dependence of the fluctuations with the dimension $D$ of the Hilbert space.  In this work, we
identify the exponent of this power-law behavior as $-\frac{1}{2}$.  Examining several
non-integrable models, we provide strong numerical evidence for $\sim{D}^{-1/2}$ behavior of EEV
fluctuations.  The ${D}^{-1/2}$ behavior generally becomes clear only at the largest sizes
accessible through full numerical diagonalization.  Our analysis therefore uses a comparison of
several sizes, at varying distances from integrability.  We use Hamiltonians designed to be tunable
between two integrable limits, and thus examine how this finite-size dependence is affected by
proximity to integrable points. As the integrable points are approached, larger sizes are required
for the ${D}^{-1/2}$ behavior to set in, and for purely integrable systems the size dependence is no
longer ${D}^{-1/2}$.

The exponent $-\frac{1}{2}$ suggests the central limit theorem, which would predict power-law
dependences if $\sigma_{\Delta{A}}$ is the average of $\mathcal{O}(D)$ random variables.  We
distinguish between two plausible mechanisms, and identify the correct explanation: The exponent
arises from the averaging over effectively random coefficients of individual eigenstates, and not
from an average over $\mathcal{O}(D)$ eigenstates in the definition of $\sigma_{\Delta{A}}$.  This
explanation relies on assumptions of effective randomness which are difficult to prove rigorously,
but are in the same spirit as the ETH itself.  A particularly nontrivial aspect is that it is not
immediately obvious why this argument should break down for integrable systems.  While the concept
of effective randomness provides useful insight, the unavoidably heuristic nature of such arguments
means that our numerical analysis is essential for determining the finite-size scaling of EEV
fluctuations.

We use several observables for each model Hamiltonian, to show the validity of the $D^{-1/2}$ law
for a wide variety of observables.  Unlike some of the previous studies of the ETH (e.g.,
\cite{RigolEA2008}), we do not refer to particular quench protocols, which corresponds loosely to
focusing on particular parts of the eigenspectrum.  Instead, we examine the complete spectrum, and
thus a broad class of quantum quenches.  The robustness of our results for different observables,
Hamiltonians, and quench protocols, provides compelling evidence for the universality of the
$D^{-1/2}$ scaling.

The structure of this article is as follows. In Sec.~\ref{sec_eev}, we introduce our measure for the
amplitude of EEV fluctuations. We define our models and observables in
Sec.~\ref{sec_models_and_observables}. The $D^{-1/2}$ scaling of the EEV fluctuations is presented
in Sec.~\ref{sec_scaling}, where we give both numerical results and a heuristic argument. The
conclusion and discussion appears in Sec.~\ref{sec_conclusion}. 
The Appendices provide further details: App.~\ref{sec_app_microcanonical} discusses issues related
to our definition of the EEV fluctuations, App.~\ref{sec_app_eevfluct_decay} elaborates on the
heuristic argument for $D^{-1/2}$ scaling, and App.~\ref{sec_app_sparse_methods} provides detail on
the numerical methods.

\section{Formulation; EEV fluctuations}
\label{sec_eev}%
The ETH states that the diagonal matrix element of a typical operator $\hat{A}$ in the eigenstates
$\ket{\psi_\alpha}$ of the Hamiltonian, i.e., the EEVs
$A_{\alpha\alpha}=\bramidket{\psi_\alpha}{\hat{A}}{\psi_\alpha}$, vary smoothly with the
corresponding energy eigenvalues $E_{\alpha}$. Thus, the EEVs may be considered as constant within
an energy window $[E-\Delta{E},E+\Delta{E}]$.  In other words, the values of $A_{\alpha\alpha}$
approximately coincide with the microcanonical average $\avgmicro{\hat A}(E_\alpha,\Delta E)$,
defined as the average EEV within this window:
\begin{equation}\label{eqn_mca}
  \avgmicro{\hat A}(E,\Delta E)
  =\frac{1}{N_{E,\Delta E}}\sum_{\alpha:E_\alpha\in[E-\Delta E,E+\Delta E]}A_{\alpha\alpha},
\end{equation} 
where $N_{E,\Delta E}$ is the number of states in this window.  If the initial non-equilibrium
state has weights constrained to such a ``microcanonical'' window, then the ETH guarantees that the
long-time average will be equal to the canonical expectation value.

\begin{figure}[t]
  \center\includegraphics[width=0.99\columnwidth]{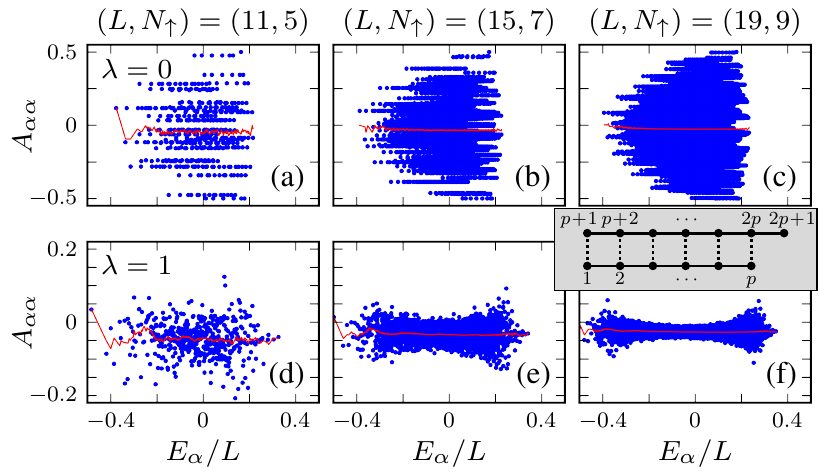}
  \caption{ 
Eigenstate expectation values (blue dots) and the microcanonical average (red curves) for the observable $\hat{A}=S^z_2$ with $\lambda=0$ (a--c) and $\lambda=1$ (d--f) for three different system sizes $L$. The energies on the horizontal axis are scaled by system size for meaningful comparison.
(The microcanonical average is nearly constant in this case, which is not representative for all
observables, see e.g., Refs.~\cite{Rigol2009PRL,Roux2010PRA81}.)
Inset: Geometry of the ladder system and site labeling.  Solid and dashed lines are
$H_{\mathrm{leg}}$ and $H_{\mathrm{rung}}$ couplings.
}
\label{fig_ladder_eev}
\end{figure}

We wish to study how this behavior is approached with increasing system size.  Therefore we study
the fluctuations around the microcanonical average as a function of size.  For every $\alpha$, we
define $\Delta{A}_{\alpha}=A_{\alpha\alpha}-\avgmicro{A}(E_\alpha,\Delta E)$.  We then consider the
statistical properties of $\Delta{A}_{\alpha}$ over a large part of the Hilbert space. In the
following, we take averages over all states in the central $20\%$ of the total energy range of the
spectrum, which we denote by $\avgcenter{\cdots}$. This average typically includes more than half
of all eigenstates. The highest and lowest end of the spectrum are left out because the spectrum
edges are likely to show atypical behavior, cf.\ Fig.\ \ref{fig_ladder_eev}.  The EEV fluctuations
$\sigma_{\Delta{A}}$ are defined as the standard deviation of $\Delta A$,
\begin{equation}\label{eqn_fluct}
  \sigma_{\Delta{A}}^2 (\Delta E)
  \equiv \bigl\langle [\Delta A_{\alpha}]^2\bigr\rangle_\mathrm{c}
  \equiv \bigl\langle [A_{\alpha\alpha}-\avgmicro{A}(E_\alpha,\Delta E)]^2\bigr\rangle_\mathrm{c}.
\end{equation}
We note that $\sigma_{\Delta{A}}$ cannot be interpreted as a standard deviation of the
$A_{\alpha\alpha}$, because of the microcanonical average $\avgmicro{\cdots}$ rather than the
ordinary average $\avg{\cdots}$ on the right-hand side of \eqn\eqref{eqn_fluct}. 
In the definition above we have assumed that the average of $\Delta A$ is
negligible, i.e., that
\begin{equation}\label{eqn_fluct_stdev_approximation}
  \mathop{\mathrm{var}}(\Delta A_{\alpha}) \equiv \avg{\Delta A_\alpha^2}_\mathrm{c} - \avg{\Delta
  A_\alpha}_\mathrm{c}^2\approx \avg{\Delta A_\alpha^2}_\mathrm{c}.
\end{equation}
While the smallness of
$\Delta{A_\alpha}$ is intuitively reasonable, the definition of $\Delta A_\alpha$ in terms of the
microcanonical average does not guarantee \emph{a priori} that 
\begin{equation}\label{eqn_fluct_stdev_condition}
\avg{\Delta A_\alpha}_\mathrm{c}^2 \, \ll\, \avg{\Delta A_\alpha^2}_\mathrm{c}
\end{equation}
is valid. Numerical evidence for the validity of this inequality is presented in
App.~\ref{sec_app_microcanonical}. Given that this condition holds, the interpretation of
$\sigma_{\Delta A}$ defined in \eqn\eqref{eqn_fluct} as a standard deviation of $\Delta A$ is
justified.

In Fig.~\ref{fig_ladder_eev},
the energies are divided by the system size $L$.  The reason is that the upper and lower parts
of the energy spectrum scale as $L$, thus the spectrum appears in the same range of $E_{\alpha}/L$.
The microcanonical average curves for the EEVs also look roughly similar for different sizes, when
plotted against $E_{\alpha}/L$, as does the density of states.

The dependence of $\sigma_{\Delta A}$ on the width $\Delta E$ of the microcanonical window is weak,
as long as the range $[E-\Delta E,E+\Delta E]$ contains sufficiently many states for good statistics
while it remains sufficiently narrow so that the microcanonical average follows the EEVs well.  As a
good tradeoff for satisfying both these conditions, we have used the value $\Delta E = 0.025L$ for
all following results. Justification for this value can be found in
App.~\ref{sec_app_microcanonical}.
As with the horizontal axes in Fig.~\ref{fig_ladder_eev}, we rescale the window width by keeping
$\Delta E /L$ constant. This window thus contains approximately equal fractions of the total number
of eigenstates for different values of $L$.

\section{Models and observables}
\label{sec_models_and_observables}%
\subsection{Tunable model Hamiltonians}
We will present results for three families of Hamiltonians.  These are designed to be tunable toward
or away from integrable limits, to have good thermodynamic limits, and to avoid symmetries that lead
to degeneracies in the spectrum.  We use systems with a Hamiltonian of the form $H=H_0+\lambda H_1$,
such that the model is integrable if the control parameter $\lambda$ is $0$ or $\infty$. For
$\lambda\in(0,\infty)$, the system is non-integrable.

The first two are based on the spin-$\tfrac{1}{2}$ anisotropic Heisenberg (XXZ) chain, which is
integrable via the Bethe ansatz \cite{fradkin1997field,*Kasteleijn1952,*Bethe1931,
  *Takahashi01121972}.  These Hamiltonians commute with the total $z$-component of spin, so that the
number $\nup$ of `up' spins is conserved.  We examine finite-size scaling by increasing $(L,\nup)$.
To suppress unwanted symmetries, e.g., $\mathrm{SU}(2)$, we take the anisotropy $\Delta$ to be away
from $0$ or $1$; results are presented for $\Delta=0.8$.

The \emph{Heisenberg ladder} consists of two coupled XXZ chains (see Fig.~\ref{fig_ladder_eev}, inset). The Hamiltonian for the $L=(2p+1)$-site model is given by $H_\mathrm{ladder}=H_\mathrm{leg}+\lambda
H_\mathrm{rung}$, where
\begin{equation}
  H_\mathrm{leg}=\sum_{i=1}^{p-1}h_{i,i+1}+\sum_{i=p+1}^{2p}h_{i,i+1}
  \ \ \text{and}\ \ %
  H_\mathrm{rung}=\sum_{i=1}^{p}h_{i,i+p}
\end{equation}
are the intrachain and interchain (rung) couplings,
respectively, given in terms of the Heisenberg XXZ coupling
\begin{equation}\label{eqn_hxxz}
  h_{i,j}
  \equiv\tfrac{1}{2}(S^+_{i}S^-_{j} + S^-_{i} S^+_{j}) + \Delta S^z_{i}S^z_{j},
\end{equation}
where $S^\pm_i = S^x_i\pm S^y_i$ and $S^z_i$ are the spin operators on site $i$ ($\hbar\equiv1$).
In order to suppress reflection symmetries, one leg has an extra site compared to the other. We will focus on the $S^z_\mathrm{total}$ sector $\nup=p$.

The second Hamiltonian is the \emph{XXZ chain in a harmonic magnetic trap}, $H_\mathrm{trap} =
H_\mathrm{XXZ}+\lambda H_\mathrm{magn}$, with the open-XXZ-chain and magnetic-field terms,
\begin{equation}\label{eqn_h_xxztrap}
  H_\mathrm{XXZ} = \sum_{i=1}^{L-1}h_{i,i+1}
  \ \ \text{and}\ \ %
  H_{\mathrm{magn}} \equiv -\sum_{i=1}^{L} B_iS^z_i,
\end{equation}
respectively. Here, $\lambda B_i$ denotes the magnetic field at site $i$ where $\lambda$ parametrizes
the strength of the trap, and $B_i = [2/(L-1)^2][i-i_0]^2$.  Here the trap
center is near the midpoint of the chain, $i_0= \tfrac{1}{2}(L+1)-\Delta{i}$, with a  
shift $\Delta{i}$ that we choose to be irrational to avoid symmetries.  The factor $2/(L-1)^2$ ensures a
meaningful thermodynamic limit.
We use the sector of filling factor $\tfrac{1}{3}$ by defining $L=3\nup$.
A harmonic trap is a particularly important manner of breaking integrability, since the classic experiment exploring the role of integrability in time evolution \cite{KinoshitaEA2006} involved dynamics in a harmonic trap.

The third Hamiltonian is the \emph{Bose-Hubbard model on an open chain},
\begin{equation}\label{eqn_hbosehubbard}
  H_\mathrm{BH} =
   -\sum_{i=1}^{L-1}(b_{i}^{\dagger}b_{i+1}+b_{i+1}^{\dagger}b_{i}) +
   \lambda\sum_{i=1}^{L}b_{i}^{\dagger}b_{i}^{\dagger}b_{i}b_{i},
\end{equation}
where $b^\dagger_i$ and $b_i$ are
the bosonic creation and annihilation operators at sites $i$ \cite{FisherEA1989,
  *JakschBruderCiracGardinerZoller_PRL98, *JakschZoller_AnnPhys2005, *lewenstein2012ultracold}. The model is integrable when
only kinetic or only interaction terms are present, i.e., in the $\lambda=0$ and $\lambda\to\infty$
limits.  We avoid reflection symmetry by modifying the interaction at site $1$ to be $1.1\lambda$
instead of $\lambda$.  We present results for half filling, i.e., the number of bosons is
$\nb=\tfrac{1}{2}L$.

\subsection{Observables}

An important issue in ETH studies is the question of which observables the ETH applies to.  To show that our main result ($\sigma_{\Delta{A}}\sim{D}^{-1/2}$  behavior) is valid for a wide range of observables, we use a number of different one-site and two-site observables.  For the ladder model,
we use the spin $z$-component $S^z_i$ at site $i$ and sums of these quantities with $i$ running over multiple sites, e.g., all sites of the bottom leg $S^z_{\textrm{bottom}}$. We also consider the two-site operators $C^z_{i,j}\equiv S^z_iS^z_j$, and sums of such operators over regions of the system. We similarly study a set of one- and two-site operators and their sums over regions of the system for the XXZ chain in a trap and for the Bose-Hubbard model: For the XXZ chain, we consider one-site (e.g., $S^z_i$) and two-site spin operators (e.g., $C^z_{i,j}= S^z_iS^z_j$ and $C^{xy}_{i,j}\equiv S^+_iS^-_j+S^-_iS^+_j$) and their sums over the middle one-third sites ($S^z_\mathrm{middle}$, $C^z_\mathrm{middle}$, and $C^{xy}_\mathrm{middle}$.  For the Bose-Hubbard model, we use on-site occupancies $n_i = b_i^{\dagger}b_i$, occupancies summed over the central sites [$n_{\mathrm{middle}} = \sum_{i=i'+1}^{L-i'}n_i/(L-2i')$ with $i'=\lfloor(L+2)/4\rfloor$], and the operators for nearest neighbor two-point and four-point correlators ($b_i^{\dagger}b_{i+1}+b_{i+1}^{\dagger}b_{i}$ and $n_i{}n_{i+1}$).
%

\begin{figure}
\centering \includegraphics{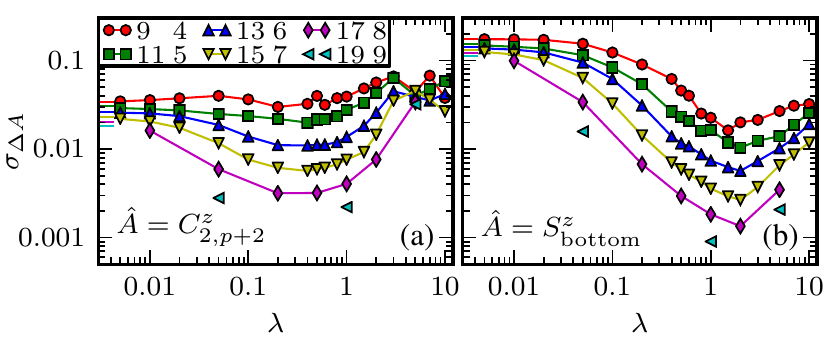}
\caption{Fluctuation amplitudes $\sigma_{\Delta A}$ as a function
  of the control parameter $\lambda$.  The $\sigma_{\Delta A}$ are presented for the Heisenberg XXZ
  ladder system for several system sizes $(L,\nup)$ (see legend) and two different observables [in
    (a) and (b), respectively].
}
\label{fig_sigma_vs_lambda} 
\end{figure}

\section{Scaling analysis of EEV fluctuations}
\label{sec_scaling}

\subsection{Dependence on size and integrability}

Figures \ref{fig_ladder_eev} and \ref{fig_sigma_vs_lambda} provide visual displays of some of the
more dramatic aspects of the ETH.

In Figure \ref{fig_ladder_eev}, we use as observable $S_2^z$, the $z$ component of the spin at site
$i=2$.  At the integrable point, the width of the distribution of EEVs can be seen to stay unchanged
with system size (top row).  For the non-integrable model, the EEV fluctuations clearly decrease
with system size in the bulk of the spectrum.  The top and the bottom of the spectrum do not show a
similarly dramatic decrease with system size, demonstrating that the ETH should be considered
relevant primarily to the bulk of the spectrum.  The physical reason is that the edges of the
spectrum tend to show emergent integrable (e.g., Luttinger liquid) behavior.

Figure~\ref{fig_sigma_vs_lambda} shows the typical dependence of EEV fluctuations on the parameter
$\lambda$ for different system sizes. This plot corroborates the intuition that for increasing
system size the fluctuations decay faster away from the integrable limits than close to them.  For
larger systems there is a pronounced minimum of  $\sigma_{\Delta{A}}$ at intermediate $\lambda$,
where it is farthest from both integrable lmiits.  

For the observable $\hat{A}=S^z_\mathrm{bottom}$ in the ladder system,
Fig.~\ref{fig_sigma_vs_lambda}(b), $\sigma_{\Delta{A}}$ is smaller in the $\lambda\sim 1$ regime
than it is in the integrable regions, even for the smallest system sizes.  For the observable
$\hat{A}=C^z_{2,p+2}=S^z_2S^z_{p+2}$, Fig.~\ref{fig_sigma_vs_lambda}(a), some deviation is seen
for very small systems, but the characteristic behavior sets in already at moderate sizes.  This
overall qualitative behavior is very typical, and is similar for all observables and all models we
have investigated.  The system size at which the crossover to large-system behavior (pronounced
minimum in the non-integrable regime) takes place depends on the model and on the observable.

\begin{figure}[t]
  \center\includegraphics[width=0.99\columnwidth]{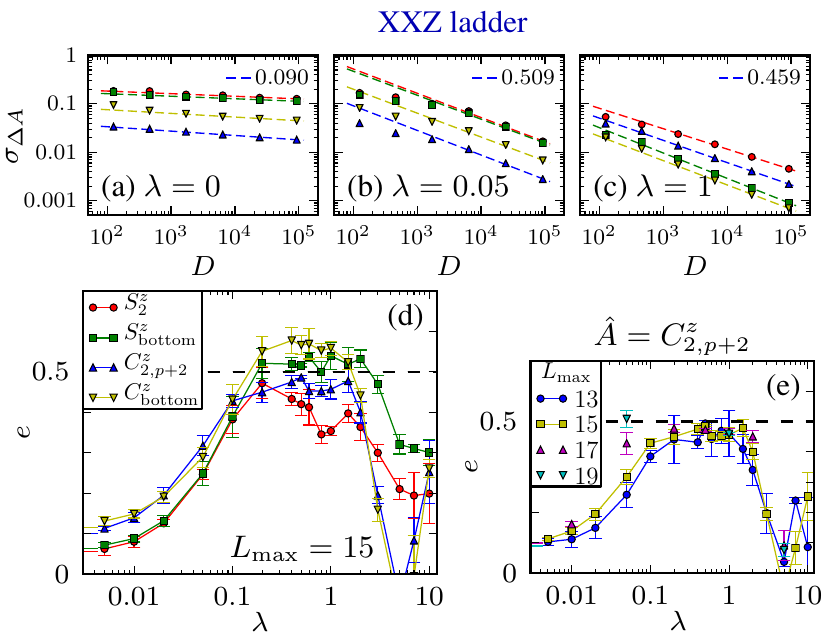}
  \caption{  \label{fig_ladder_sigma_exponent}
(a--c) Dependence of $\sigma_{\Delta{A}}$ on the Hilbert-space dimension $D$ in the
    Heisenberg ladder.  The different operators $\hat{A}$ used are indicated in the legend for panel (d).  
    The lines are best fits to $c_0D^{-e}$; the exponent estimator $e$ is indicated for the operator
    $C^z_{2,p+2}$.  
(d) The exponent estimator $e$ is plotted against $\lambda$ for the four operators.  
(e) Sizes up to $L_{\mathrm{max}}$ are used for fitting to obtain the estimator $e$.  The trend is that
    $e\to\frac{1}{2}$ for increasing $L_{\mathrm{max}}$.
}
\end{figure}

\subsection{Scaling with system size}

In Fig.~\ref{fig_ladder_sigma_exponent}(a-c), we show the dependence of the EEV fluctuations on
Hilbert-space size $D$, for several values of the rung coupling parameter $\lambda$ that tunes the
system away from integrability. The data plotted in this figure involves vertical slices of the plots in Fig.~\ref{fig_sigma_vs_lambda} (size-dependence at constant $\lambda$ values). The ETH fluctuations are commonly claimed to decrease exponentially
with system size for non-integrable models, and hence should decrease as a power law with $D$.  We
define an ``exponent'' $e$ as the one that is obtained in a power-law fit, $\sim{D}^{-e}$, to the
data for $\sigma_{\Delta A}$ for the available sizes.  We make no \emph{a priori} claims about the
dependence being actually a power law, or the obtained values of $e$ being the actual exponent in
the large-size limit.  In cases where the dependence is a power law, as expected in non-integrable
systems, $e$ is an estimator for the actual exponent.  The exponent estimator $e$ goes toward zero
as one approaches the integrable points $\lambda=0$ or $\lambda=\infty$.  At the point $\lambda=0$
the dependence on $D$ is presumably not even a power law.

\begin{figure}[t]
  \center\includegraphics[width=0.99\columnwidth]{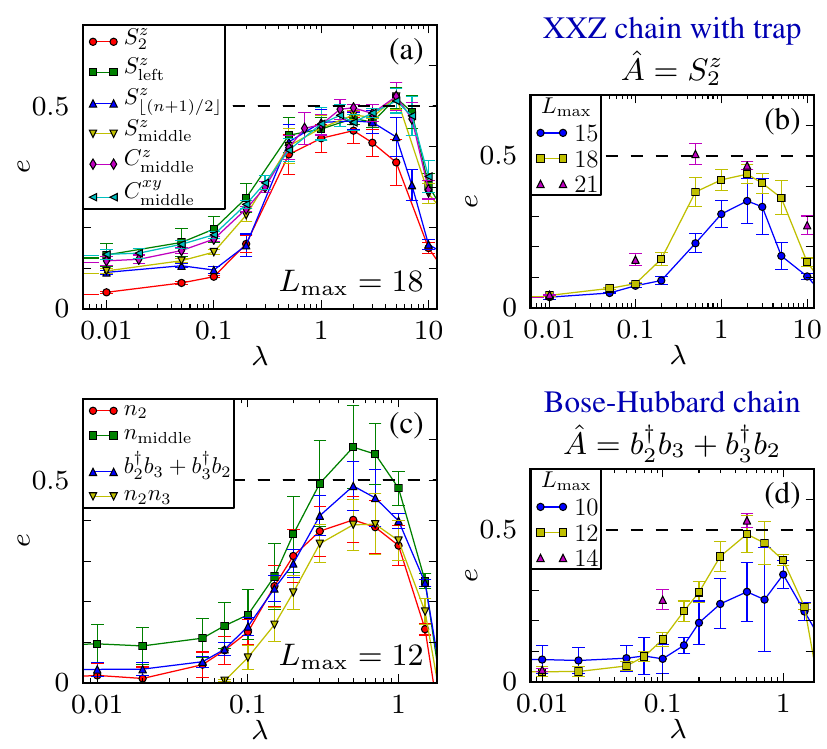}
  \caption{
For the XXZ chain in a trap, we plot the exponent estimator $e$ against integrability-breaking parameter $\lambda$ in (a). The estimator $e$ calculated with data up to $L_{\mathrm{max}}=18$. The estimator $e$ is calculated with data up to $L_{\mathrm{max}}$ sites. In (b), we plot the estimator $e$ obtained from fits with increasing $L_{\mathrm{max}}$.
Analogous results for the Bose Hubbard chain are shown in (c) (with $L_{\mathrm{max}}=12$) and (d).
}
\label{fig_trap_bh_exponent}
\end{figure}

Values of the exponent estimator $e$ are plotted in Fig.\ \ref{fig_ladder_sigma_exponent}(d,e) for
the ladder system and in Fig.\ \ref{fig_trap_bh_exponent} for the XXZ-trap system and the
Bose-Hubbard system.  There is a clear and general trend for $e$ to cluster around or approach $0.5$
in all systems, when away from integrability.  Taken together, we believe this provides compelling
evidence of $\sigma_{\Delta{A}}\sim{D}^{-1/2}$ dependence in generic non-integrable systems for
generic few-body observables $\hat{A}$.  Figure \ref{fig_ladder_sigma_exponent}(d) displays the
general behavior for several different observables in the XXZ ladder: $e\approx\tfrac{1}{2}$ for intermediate
$\lambda$ and vanishing $e$ for $\lambda$ approaching $0$ or $\infty$. Similar behavior is observed
for the XXZ chain with a trap and for the Bose-Hubbard chain, see
Figs.\ \ref{fig_trap_bh_exponent}(a) and (c), respectively.  For the three systems considered here,
the results are qualitatively similar. The general trend is that for a fixed maximum system size
$L_\mathrm{max}$, the exponent estimator $e$ clusters around $\tfrac{1}{2}$ for intermediate values
of $\lambda$, and has lower values close to the integrable limits.
In Figs.\ \ref{fig_ladder_sigma_exponent}(e), \ref{fig_trap_bh_exponent}(b) and \ref{fig_trap_bh_exponent}(d), we show more
quantitative scaling behavior for the three tunable Hamiltonians, by plotting the exponent
estimators $e$ derived from power-law fits to the data of system sizes up to $L_{\mathrm{max}}$. We note a crossover from integrable-like to $\sim D^{-1/2}$ behavior for non-integrable systems close to integrability [e.g., for the XXZ ladder at $\lambda=0.05$; see Figs.~\ref{fig_ladder_sigma_exponent}(b) and (e)], as the system size is increased. The
trend with increasing $L_{\mathrm{max}}$ points to the large-system behavior of $\sigma_{\Delta{A}}$
being $\sim{D}^{-1/2}$ over the full range $\lambda\in(0,\infty)$.

\subsection{ $D^{-1/2}$ behavior from eigenstate size: A heuristic argument}
\label{subsec_heuristic}

The $D^{-1/2}$ dependence of $\sigma_{\Delta{A}}$ can be argued heuristically by considering
projections of eigenstates onto the eigenbasis of the $\hat{A}$ operator, and then invoking the
central limit theorem.  If $\{a_{\gamma}\}$ and $\{\ket{\phi_{\gamma}}\}$ denote the eigenvalues and
eigenvectors of $\hat{A}$, and we expand the eigenvectors $\ket{\psi_\alpha}$ of $H$ as
\begin{equation}\label{eqn_a_expansion}
  \ket{\psi_{\alpha}} =
\sum_{\gamma}c_\gamma^{(\alpha)}\ket{\phi_{\gamma}}
\end{equation}
then we can write the EEVs as
\begin{equation}\label{eqn_eev_expansion}
  A_{\alpha\alpha}
   = \sum_{\gamma=1}^{D} \abs{c^{(\alpha)}_\gamma}^2 a_\gamma.
 \end{equation}
This is an average of $X_{\gamma} \equiv D\abs{c_{\gamma}^{(\alpha)}}^2a_{\gamma}$.  Under the
hypothesis that the $X_{\gamma}$ can be regarded as random variables with $D$-independent variance,
the central limit theorem guarantees that the fluctuations of $A_{\alpha\alpha}$ decrease as
$D^{-1/2}$.  (This argument is detailed further in App.~\ref{sec_app_eevfluct_decay}.)
We are unable to \emph{prove} the idea that $X_{\gamma}$ or $c_{\gamma}$ act as random variables.
However, one can intuitively think of an eigenstate of a non-integrable system as being so complex
that its projections onto the eigenbasis of a typical observable are effectively random.  This is
similar in spirit to the ETH itself (also difficult to prove rigorously), for which the argument is
that when eigenfunctions are complex enough, EEVs of typical observables will contain no signature
of the detailed structure of the wavefunction.

\begin{figure}[t]
  \center\includegraphics[width=1.0\columnwidth]{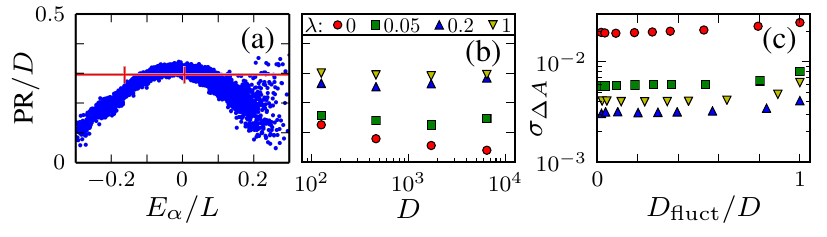}
  \caption{(a) Scaled participation ratios (PR) of
    all eigenstates with respect to the computational basis for $15$-site ladder, with
    $\lambda=1$. 
The averaging region is shown by vertical bars and the average PR value is shown by the horizontal
line.   
(b) Dependence of the scaled average PR on $D$.  
    (c) Dependence of $\sigma_{\Delta A}$ on the number $D_\mathrm{fluct}$ of states used as
    input to compute this quantity.  Data for $17$-site ladder, $\hat{A}=C^z_{2,p+2}$ for four
    different values of $\lambda$ [captions in (b)].}
  \label{fig_PR}
\end{figure}

This argument relies on the assumption that the size (number of components) of the individual
eigenfunctions is $\mathcal{O}(D)$.  This is justified in Fig.\ \ref{fig_PR}(a,b) through the
participation ratio (PR) in the computational (site) basis, defined for each eigenstate as
\begin{equation}\label{eqn_pr}
  P_\alpha=\left[\sum_\gamma\abs{c_{\gamma}^{(\alpha)}}^4\right]^{-1}.
\end{equation}
The PR measures the number of
basis states contributing to the eigenstate.  
[For a single state, the commonly discussed inverse participation ratio (IPR) is $1/P_\alpha$.]
Figure~\ref{fig_PR}(b) shows that it is, on average,
indeed proportional to $D$ in non-integrable cases.
The $D^{-1/2}$ scaling of $\sigma_{\Delta A}$ and the $D$ scaling of the average PR are, taken
together, consistent with the expectation \cite{NeuenhahnMarquardt2012} that $\sigma_{\Delta A}^2$
should be proportional to the average inverse PR.  
The observations of Ref.~\cite{NeuenhahnMarquardt2012}, in terms of the average inverse PR in the
momentum Fock basis, can also be cast as a heuristic argument for $D^{-1/2}$ scaling, roughly
equivalent to the reasoning above.

We emphasize that the number $D_\mathrm{fluct}$ of states included in the average
$\avgcenter{\cdots}$ does not account for the $D^{-1/2}$ dependence. The quantity
$\sigma_{\Delta{A}}$ is the standard deviation of the distribution of the $\Delta A_\alpha$'s and it
is independent of how many times one ``probes'' this distribution, i.e., the number of states that
is used to compute $\sigma_{\Delta A}$. As shown in Fig.~\ref{fig_PR}(c), the value of
$\sigma_{\Delta A}$ is independent of $D_\mathrm{fluct}$ as long as states at the edges of the
spectrum are avoided. The slight dependence on
$D_\mathrm{fluct}$ is caused by the fact that at the edges of the spectrum, the fluctuations of
$A_{\alpha\alpha}$ are different from those in the center of the spectrum.  The edges of the
spectrum, of course, are outside the purview of the ETH.  This demonstrates that the $D^{-1/2}$
behavior arises not from the number of eigenstates averaged over, but from the complexity of the
individual eigenstates themselves.

The fluctuations decrease more slowly with system size at the integrable points, as evidenced by the
vanishing of the exponent estimator $e$ in the $\lambda\to0$, $\infty$ limits in each of the
models. This implies a difference in the structure of the individual eigenstates.  One
characterization of this difference is visible in Fig.\ \ref{fig_PR}(b) where the scaled average
PR for the integrable model is seen to decrease with system size.  A detailed study of eigenstates
in integrable models from this perspective, to complement the studies of
Refs.\ \cite{CassidyEA2011,RigolFitzpatrick2011,HeRigol2012,GramschRigol2012,KollarEA2011,HeEA2013,SteinigewegEA2013,IkedaEA2013},
is interesting but is beyond the scope of the present work.

\section{Summary \& Discussion}
\label{sec_conclusion}

For non-integrable systems, we have presented the size dependence of the deviation from ETH, as
measured by the EEV fluctuations, $\sigma_{\Delta{A}}$, for lattice systems.  It is well accepted
that $\sigma_{\Delta{A}}$ decreases exponentially with the system size,
$\sigma_{\Delta{A}}\sim\exp[-c_1L]$, e.g., Ref.\ \cite{SteinigewegEA2013} has numerical data showing
the exponential decay.  In terms of the Hilbert-space size $D$, if $D\sim\alpha^L$ (see
App.~\ref{subsec_app_hilbert_sparse_sizes}), 
then
\begin{equation}
  \sigma_{\Delta{A}}\sim{D}^{-e}\sim\exp[-e(\ln\alpha)L] .
\end{equation}  
Our work makes this relationship
precise by determining the exponent to be $e=\frac{1}{2}$, or equivalently, the coefficient to be
$c_1=\frac{1}{2}\ln\alpha$.

The $D^{-1/2}$ behavior is difficult to convincingly show from calculations for a fixed
non-integrable Hamiltonian.  We have therefore used a control parameter to move away from an
integrable Hamiltonian; this makes clear the trend of $e$ approaching $\tfrac{1}{2}$ as one tunes away from
integrability.  In addition, the sizes at which the $D^{-1/2}$ behavior sets in are at the limit of
sizes that can be comfortably addressed by full numerical diagonalization, which is the method used
in current numerical studies of the ETH.  Our use of sparse matrices with shift-and-invert algorithm
(see App.~\ref{subsec_app_sparse_methods}) has allowed us to reach larger sizes: we have used full diagonalization for sizes up to
$D\sim 2\times 10^4$, and sparse matrix methods for larger $D$, the largest being above $10^5$.

While the exponent $e=\tfrac{1}{2}$ has not, to the best of our knowledge, appeared for the EEV
fluctuations in the setting of condensed-matter Hamiltonians, some similar or related results exist.
The observations of Ref.~\cite{NeuenhahnMarquardt2012} could be combined to construct an argument
for $D^{-1/2}$ scaling, as discussed in \ref{subsec_heuristic}.  In the literature on `typicality'
\cite{GemmerEA2009book, DubeyEA2012, GemmerMichel2006, Reimann2007, GoldsteinEA2006}, there is the
expectation that the deviation of random Hamiltonians from typicality (closely related to ETH)
scales with system size such that measures of atypicality behave as ${\sim}D^{-1/2}$.
Ref.~\cite{DubeyEA2012} shows this numerically for random Hamiltonians but finds other exponents for
spin-chain Hamiltonians, for the sizes treated.  Further work is needed for a full understanding
of the connection between these results and ours. 

Our work opens up several new questions.  First, the $D^{-1/2}$ behavior does not set in at smaller
sizes.  It is obvious from Fig.\ \ref{fig_ladder_sigma_exponent}(d,e) and
Fig.\ \ref{fig_trap_bh_exponent} that larger sizes are necessary when integrability is weakly
broken, since the $e$ values calculated from available sizes do not reach $\tfrac{1}{2}$ for
$\lambda$ near the integrable points.  This indicates a length scale associated with the degree of
integrability breaking, a concept that might be possible to explore quantitatively.  Second, our
quantitative result $\sigma_{\Delta{A}}\sim{D}^{-1/2}$ requires a finite Hilbert-space dimension
$D$.  It is not obvious how to generalize this law to continuum systems.  Finally, it would be
interesting to ask how the finite-size behavior of EEVs is affected by proximity to (many-body)
localization, which, like integrability, is expected to be detrimental for thermalization.

\acknowledgments
We thank P.~Ribeiro, M.~Rigol, and A.~Sen for useful discussions.

\appendix

\section{EEV fluctuations: Definition issues}
\label{sec_app_microcanonical}

\begin{figure}
  \includegraphics{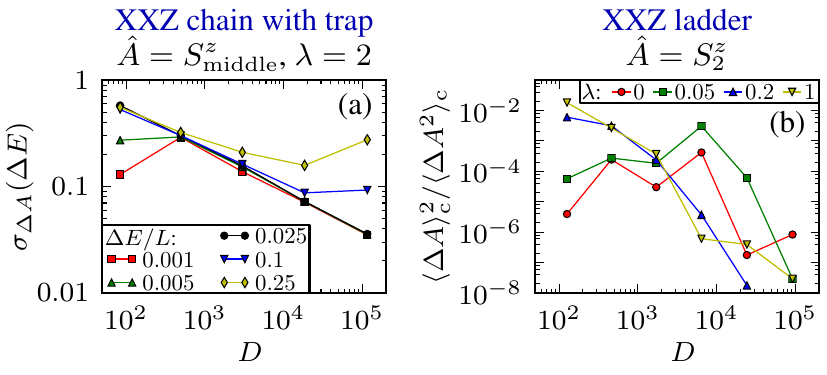}
  \caption{(a) Fluctuation amplitudes as a function of $D$ with for several widths $\Delta E$ of the microcanonical window. The value $\Delta E/L=0.025$ (black curve) is used for the analysis in the rest of this work. The values $0.001$ and $0.25$ are extreme cases. (b) Average value $\avg{\Delta A_\alpha}_\mathrm{c}^2$ divided by $\avg{\Delta A_\alpha^2}_\mathrm{c}$ as a function of Hilbert-space dimension $D$ for the ladder system with $\hat{A} = S^z_2$ for several values of $\lambda$. The lines connecting the points serve as a guide to the eye.}
  \label{fig_fluct_analysis} 
\end{figure}

The width of the microcanonical window has been chosen to satisfy $\Delta E = 0.025L$, as a good compromise between the conditions that it contain sufficiently many eigenstates for good statistics, and that the microcanonical average follows the EEVs well. To justify the choice of this value, we have plotted the fluctuation amplitudes $\sigma_{\Delta A}$ as a function of system size $D$ for several values of $\Delta E$ in Fig.~\ref{fig_fluct_analysis}(a). Here, we have chosen the observable $\hat{A}=S^z_\mathrm{middle}$ for the trap system at $\lambda=2$, which is characterised by a strongly nonlinear dependence of the microcanonical average $\avgmicro{A}(E,\Delta E)$ on $E$.
This situation is a worst-case scenario: We expect a relatively strong dependence of the resulting $\sigma_{\Delta A}$ on $\Delta E$, because for very large values, the microcanonical average does not follow the actual EEVs $A_{\alpha\alpha}$ well. This mechanism is responsible for the fact that for very large values of $\Delta E$, the fluctuations are overestimated [see Fig.~\ref{fig_fluct_analysis}(a)]. In the case where $\avgmicro{A}(E,\Delta E)$ would depend almost linearly on $\Delta E$, the dependence of $\sigma_{\Delta A}$ on $\Delta E$ will be weaker. Another feature that we find from Fig.~\ref{fig_fluct_analysis}(a) is that the fluctuations are underestimated if the number of states in the microcanonical average is very small, in the case of small $\Delta E$ and small system size. Finally, we may conclude from Fig.~\ref{fig_fluct_analysis}(a) that $\sigma_{\Delta A}$ is almost independent of $\Delta E$ for a large range of values around $\Delta E/L=0.025$. Here, we emphasize that the values $\Delta E/L=0.001$ and $\Delta E/L=0.25$ present very extreme cases, where the microcanonical window encompasses only a few eigenstates (for the smaller system sizes) and almost the whole spectrum, respectively.

In order for the interpretation of $\sigma_{\Delta A}$, as defined by \eqn\eqref{eqn_fluct}, as standard deviation of the $\Delta A_\alpha$ to be valid, we must test the condition that the average of $\Delta A_\alpha$ is negligibly small, expressed by \eqn\eqref{eqn_fluct_stdev_condition}. In Fig.~\ref{fig_fluct_analysis}(b), we plot the ratio $\avg{\Delta
  A_\alpha}_\mathrm{c}^2/\avg{\Delta A_\alpha^2}_\mathrm{c}$ for the observable $\hat{A}=S^z_2$ in
the ladder system, as a function of the Hilbert space $D$ and for several values of $\lambda$. We
indeed observe that $\avg{\Delta A_\alpha}_\mathrm{c}^2$ is negligibly small compared to
$\avg{\Delta A_\alpha^2}_\mathrm{c}$. The approximation
$\mathop{\mathrm{var}}(\Delta A_{\alpha}) \equiv \avg{\Delta A_\alpha^2}_\mathrm{c} - \avg{\Delta
  A_\alpha}_\mathrm{c}^2\approx \avg{\Delta A_\alpha^2}_\mathrm{c}$
generally improves for increasing system size. Thus, the interpretation of $\sigma_{\Delta A}$ as standard deviation of the $\Delta A_\alpha$ is justified.


\section{Mechanism for $D^{-1/2}$ decay of EEV fluctuations}  \label{sec_app_eevfluct_decay}

In this section, we expand on the argument provided in Sec.~\ref{subsec_heuristic} for the  $D^{-1/2}$ decay of EEV
fluctuations.   The  $D^{-1/2}$ behavior arises from the fact that the individual eigenstates have
$D$ components, and not from the sum over $\mathcal{O}(D)$ different eigenstates in the definition
of $\sigma_{{\Delta}A}$.

\paragraph*{$D^{-1/2}$ from randomness of coefficients ---}

Our argument is based on the expansion of the energy eigenstates $\ket{\psi_\alpha}$ in the basis of
eigenvectors $\ket{\phi_{\gamma}}$ (with eigenvalues $a_\gamma$) of the operator $\hat{A}$, as given by \eqn\eqref{eqn_a_expansion}. The EEVs are then realizations of a random variable which is the
average of $D$ approximately random variables,
\begin{equation}
 A_{\alpha\alpha}
   = \sum_{\gamma=1}^{D} \abs{c^{(\alpha)}_\gamma}^2 a_\gamma
   = \frac{1}{D}\sum_{\gamma=1}^{D} X_{\gamma},
\end{equation}
where $X_{\gamma} = D \abs{c^{(\alpha)}_\gamma}^2 a_{\gamma}$.  We will now regard $X_{\gamma}$ as
random, quasi-independent, variables.  There is no rigorous justification for this, but it can be
argued in the same spirit as the arguments in favor of the ETH itself, namely, in a large
non-integrable system the typical eigenstate is so complicated that its components are effectively
random in any reasonable basis. In principle, the randomness of $\abs{c^{(\alpha)}_\gamma}^2$ and of
$X_{\gamma}$ may be different, due to the multiplication with the eigenvalues $a_\gamma$. However,
if these eigenvalues take only very few ($\ll D$) different values, then
$\abs{c^{(\alpha)}_\gamma}^2$ is random if and only $X_{\gamma}$ is.

Assuming that the $X_{\gamma}$ act as random variables, the central limit theorem implies that the
EEVs have the standard deviation $\sqrt{\mathop{\mathrm{var}}(X_{\gamma})/D}$.  If the variance of
$X_\gamma$ is approximately $D$-independent (as argued below), the $D^{-1/2}$
dependence of the fluctuations follows immediately.

We emphasize again that our reasoning is based on the assumption that the
$\abs{c^{(\alpha)}_\gamma}^2$ are ``random enough'' that the central limit theorem can be used.  The
extent or exact nature of this randomness is not understood in detail, to the best of our knowledge.
At or near integrability, $\sigma_{\Delta A}$ no longer scales as $D^{-1/2}$, which suggests that
the coefficients $\abs{c^{(\alpha)}_\gamma}^2$ lose their randomness in such situations.  

Even in the non-integrable case, the assumption is invalid for any \emph{conserved} quantity $A$. If
$\hat{A}$ commutes with $\hat{H}$, one can choose a common eigenbasis, and consequently only
one $c^{(\alpha)}_\gamma$ is nonzero.

\paragraph*{$\mathrm{var}(X_\gamma$) is independent of $D$ ---}

We now argue that the variance of $X_\gamma = D\abs{c^{(\alpha)}_\gamma}^2 a_{\gamma}$ is
independent of $D$.  The eigenvalues $a_\gamma$ of the operator $\hat{A}$ are typically polynomial
in system size, and hence at most logarithmic in $D$. In addition, the average value of
$\abs{c^{(\alpha)}_\gamma}^2$ is $1/D$ by normalization.  If the distribution of
$\abs{c^{(\alpha)}_\gamma}^2$ is not extremely pathological, this implies that the variance of
$\abs{c^{(\alpha)}_\gamma}^2$ scales as $1/D^2$.  With this observation, it follows that
$\mathop{\mathrm{var}}(X_\gamma)\sim 1$, i.e., constant in system size.

\begin{figure}
  \center\includegraphics{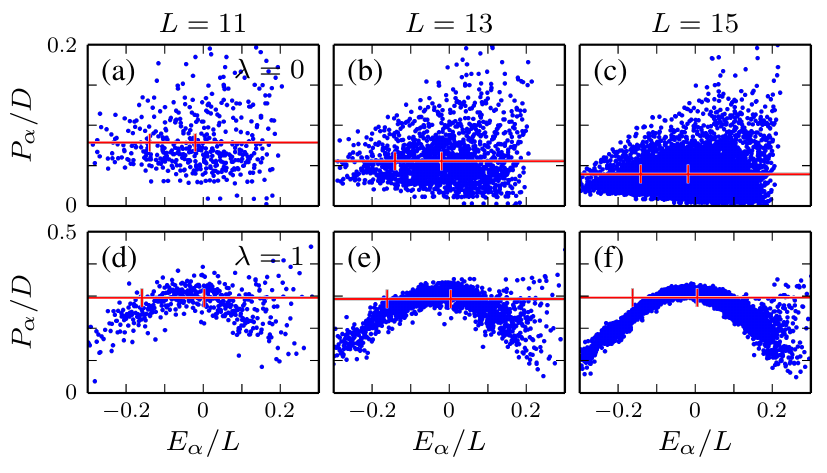}
  \caption{Scaled participation ratios $P_\alpha/D$ as a function of $E_\alpha$ in the Heisenberg
    XXZ ladder system.  We show the results for system sizes $L=11,13,15$ and two values of the
    control parameter: (a)--(c) $\lambda=0$, (d)--(f) $\lambda=1$.  The horizontal red line
    indicates the average scaled PR. The two vertical bars enclose the middle $20\%$ of the
    spectrum, the energy region for which the average is taken. Here, we have chosen the
    computational basis, which is also an eigenbasis for most of the observables discussed in this
    text (e.g., $S^z_i$, $S^z_iS^z_{i+1}$).}
  \label{fig_pr_app}
\end{figure}

The variance of $X_\gamma$ can be related to the participation ratio (PR) through
\begin{equation}
  P_\alpha/D=[1+\mathop{\mathrm{var}}(D\abs{c^{(\alpha)}_\gamma}^2)]^{-1}\sim[1+\mathop{\mathrm{var}}(X_\gamma)]^{-1},
\end{equation}
where the PR has been defined in \eqn\eqref{eqn_pr}.  Thus, our previous statements are confirmed if
the average scaled PR is constant as a function of system size. In addition to Fig.~\ref{fig_PR}, we
present a more detailed view of the PRs in Fig.~\ref{fig_pr_app}.  The average scaled PR decreases
noticeably with size in the integrable case, while it remains constant for nonintegrable systems.

\paragraph*{Normal distribution of $A_{\alpha\alpha}$ ---}

\begin{figure}
  \center\includegraphics{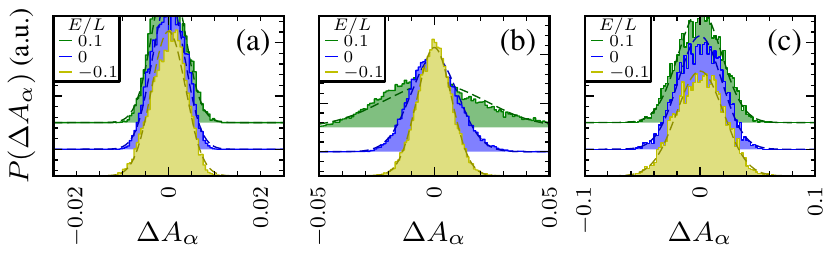}
  \caption{Histograms of the fluctuations $\Delta A_\alpha$ in a microcanonical window. (a) Ladder system, $(L,\nup)=(19,9)$, $\hat{A}=C^z_{2,p+2}$, $\lambda=1$. (b) Magnetic-trap system, $(L,\nup)=(21,7)$, $\hat{A}=S^z_2$, $\lambda=2$. (c) Bose-Hubbard system, $(L,\nb)=(14,7)$, $\hat{A}=b_2^\dagger b_3+b_3^\dagger b_2$, $\lambda=0.5$. For each system, we show the distributions at the three energies defined by $E/L=-0.1,0,0.1$. The dashed curves indicate normal distributions with the same variances as those of the fluctuations $\Delta A_\alpha$.}
  \label{fig_eevhist_app}
\end{figure}

The central limit theorem does not only give a value for the variance, it also states that the
distribution of the $A_{\alpha\alpha}$ variables should be a normal distribution for large $D$.  In
support of this statement, we present the distributions of the fluctuations $\Delta A_\alpha$ in
Fig.~\ref{fig_eevhist_app}. The distributions show the fluctuations within one window of the
microcanonical average centered at $E/L =-0.1$, $0$, and $0.1$. The results closely resemble normal
distributions, indicated by the dashed curves.  This provides indirect support to the conjecture
that the coefficients  $\abs{c^{(\alpha)}_\gamma}^2$ are effectively random.

\section{Computational details: sparse-matrix methods and Hilbert-space sizes}  \label{sec_app_sparse_methods}

Hamiltonians in condensed matter physics generally lead to sparse matrices, so that it is often
advantageous to use sparse matrix methods like the Lanczos algorithm, which accesses the lowest or
highest parts of the eigenspectra.  In studies of the ETH, however, we explicitly want to access
parts of the spectrum away from the edges.  In addition, we have taken the approach of looking at
the entire bulk of the spectrum.  Therefore, as conventional in computational research on the ETH,
we have used full diagonalization of the Hamiltonian matrix, in order to treat Hilbert-space
dimensions up to $D\approx 20000$.  However, in this work, we have additionally gone beyond this
size limit, by using sparse matrix methods that access non-extremal parts of the spectrum.  This
method is described in subsection \ref{subsec_app_sparse_methods}.  In subsection
\ref{subsec_app_hilbert_sparse_sizes} we connect Hilbert-space sizes to system sizes, for our three
model systems.

\subsection{ Sparse-matrix methods }
\label{subsec_app_sparse_methods}

In order to tackle larger systems than can be comfortably accessed with full diagonalization on
present-day machines, we have used a divide-and-conquer technique to split the problem of
diagonalization into smaller parts.  We used the so-called \emph{shift-invert} algorithm: for a
matrix $H$ and a chosen value $\gamma$, one applies Lanczos diagonalization to the matrix $(H-\gamma
I)^{-1}$, so that one effectively finds the eigenvalues of $H$ close to $\gamma$.  In practice, one
does not invert the matrix explicitly, since that would generate a non-sparse inverse matrix.
Instead, the generation of the Krylov basis $\{\psi_i\}$, defined through $\psi_{i+1} = (H-\gamma
I)^{-1}\psi_{i}$, is performed by iteratively solving $(H-\gamma I)\psi_{i+1}=\psi_{i}$.  There is
thus an ``inner'' iteration necessary for generating the Krylov basis, in addition to the usual
Lanczos iteration.  Such methods are often known as ``inner-outer'' iterative methods.

While this method clearly takes significantly more run-time than bare Lanczos diagonalization, it
has the advantage that any part of the spectrum can be accessed. For intermediate Hilbert-space
sizes ($D\sim10000$), we have performed several comparisons between the results of the sparse and
the dense method, and we have found them to yield consistent results.

In order to find all eigenvalues and eigenstates of a large sparse matrix, we choose a set of
initial energies $\{\gamma_i\}$, and compute in parallel typically $2000$ eigenvalues close to each
of these values together with the EEVs for a set of observables. Each application of the
shift-invert method yields the eigenvalues within a certain (\emph{a priori} unknown) energy
interval. Afterwards, the results are ``patched'' together, i.e., for the energy regions where two
or more such intervals overlap, the eigenvalues and EEVs are compared, and duplicates are removed
such that each appears only once in the final result. Finally, the total number of eigenvalues is
compared against the known dimension of the Hilbert space. 
If the result does not contain all the eigenstates, more shift-invert diagonalizations are performed
until all eigenstates have been obtained.
The largest system for which we have found the full eigenspectrum
using this procedure is of Hilbert-space dimension $D=116280$.

\begin{table}
  XXZ ladder; $u=2$, $v=1$, $w=1$:\\
  \begin{tabular}{l|rrrrrr}
    $p$   & $4$ & $5$ & $6$ & $7$ & $\mathbf{8}$ & $\mathbf{9}$\\
   \hline
    $L$   & $9$ & $11$& $13$& $15$& $\mathbf{17}$& $\mathbf{19}$\\
    $\nup$& $4$ & $5$ & $6$ & $7$ & $\mathbf{8}$ & $\mathbf{9}$\\
    \hline
    $D$   & $126$ & $462$ & $1716$ & $6435$ & $\mathbf{24310}$ & $\mathbf{92378}$
  \end{tabular}\\[2ex]
  XXZ trap; $u=3$, $v=1$, $w=0$:\\
  \begin{tabular}{l|rrrrr}
    $p$   & $3$ & $4$ & $5$ & $6$ & $\mathbf{7}$ \\
   \hline
    $L$   & $9$& $12$& $15$& $18$& $\mathbf{21}$\\
    $\nup$& $3$ & $4$ & $5$ & $6$ & $\mathbf{7}$\\
    \hline
    $D$   & $84$ & $495$ & $3003$ & $18564$ & $\mathbf{116280}$
  \end{tabular}\\[2ex]
  Bose-Hubbard; $u=2$, $v=1$, $w=0$:\\
  \begin{tabular}{l|rrrrr}
    $p$   & $3$ & $4$ & $5$ & $6$ & $\mathbf{7}$ \\
   \hline
    $L$   & $6$ & $8$& $10$& $12$& $\mathbf{14}$\\
    $\nb$& $3$ & $4$ & $5$ & $6$ & $\mathbf{7}$\\
    \hline
    $D$   & $56$ & $330$ & $2002$ & $12376$ & $\mathbf{77520}$
  \end{tabular}
  \caption{  \label{tbl_app_systemsize}
Overview of the system sizes $L$ and Hilbert-space dimensions $D$ of the models used. The cases with
bold-faced values have been investigated with the sparse diagonalization algorithm; in all other
cases, full diagonalization has been used. }
\end{table}

\subsection{Hilbert-space dimensions and system sizes}
\label{subsec_app_hilbert_sparse_sizes}

Our results on the EEV fluctuations have been presented in terms of the Hilbert-space dimension
$D$. In order to ``translate'' the results to system size $L$, one uses the relations
\begin{equation}\label{eqn_app_dim}
  D=\binom{L}{\nup}
  \qquad\text{and}\qquad
  D=\binom{L+\nb-1}{\nb}
\end{equation}
for an $L$-site XXZ model with $\nup$ spins up and for an $L$-site Bose-Hubbard model with $\nb$ bosons, respectively.
In our numerical study, we approach the thermodynamic limit with systems with (almost) constant filling fraction $f\equiv v/u$ for integers $u$ and $v$. We perform our calculations for the sequences  $(L,\nup)=(u p+w,vp)$ and $(L,\nb)=(u p+w,vp)$ ($p=1,2,\ldots$) for the XXZ and Bose-Hubbard models, respectively; $w$ is an additional constant integer. Table~\ref{tbl_app_systemsize} provides an overview of the choices of the parameters and the resulting system and Hilbert-space sizes for the models discussed in this work.

With a constant filling fraction $v/u$, the Hilbert-space dimensions of \eqn\eqref{eqn_app_dim} can be approximated using Stirling's formula, as
\begin{equation}\label{eqn_app_dimratio}
  D \to \frac{\sqrt{c_{u,v}}}{\sqrt{2\pi p}}(\beta_{u,v})^p,
\end{equation}
where $c_{u,v}$ equals $u/v(u-v)$ for the XXZ models and $(u+v)/uv$ for the Bose-Hubbard model, and
\begin{equation}
  \beta_{u,v}\equiv\left\{\begin{array}{ll}
                            {u^u}/{v^v (u-v)^{u-v}}\quad&\mbox{(XXZ)}\\
                            {(u+v)^{u+v}}/{u^u v^v}\quad&\mbox{(Bose-Hubbard)}
  \end{array}\right.
\end{equation}
defines the limiting ratio $\lim_{p\to\infty}D_{p+1}/D_{p}$ between the Hilbert-space dimensions of two subsequent realizations in the sequence of system sizes. In other words, the dimension of the Hilbert space is approximately exponential in the system size $L$, as $D\sim L^{\beta_{u,v}}$. 

Assuming the power-law behavior $\sigma_{\Delta A} \propto D^{-e}$ of the EEV fluctuations (with $e=\tfrac{1}{2}$ for non-integrable models), we find that this quantity scales exponentially in the system size, as $\sigma_{\Delta A} \approx \mathrm{const} \times (2\pi L)^{e/2}(\zeta_{f})^{-eL}$, where
\begin{equation}
  \zeta_f\equiv(\beta_{u,v})^{1/u}
  =\left\{\begin{array}{ll}
            1/f^f(1-f)^{1-f}\quad&\mbox{(XXZ)}\\
            (1+f)^{1+f}/f^f\quad&\mbox{(Bose-Hubbard)}
          \end{array}\right.,
\end{equation}
in terms of the filling fraction $f$.



%

\end{document}